\documentclass[runningheads]{llncs}
\usepackage{graphicx}
\usepackage{latexsym}

\usepackage{multirow}%
\usepackage{amsmath,amssymb,amsfonts}%
\usepackage{mathrsfs}%
\usepackage[title]{appendix}%
\usepackage{xcolor}%
\usepackage{textcomp}%
\usepackage{manyfoot}%
\usepackage{booktabs}%
\usepackage{algorithm}%
\usepackage{algorithmicx}%
\usepackage{algpseudocode}%
\usepackage{listings}%
\usepackage{subcaption}
\usepackage{caption}
\usepackage{comment}
\usepackage{longtable}
\usepackage{listings}
\usepackage{xcolor} % for colors

\def\urltilda{\kern -.15em\lower .7ex\hbox{\~{}}\kern .04em}

\newcommand{\Reject}{{\mathit{REJECT}}}

\floatname{algorithm}{Procedure}

\begin{document}

\pagestyle{headings}
%In order to omit page numbers and running heads
%please change this line to
%\pagestyle{empty}
%and change the first command line too, see above.

\mainmatter

\title{Structured Program Synthesis using LLMs: Results and Insights from the IPARC Challenge}
\titlerunning{Structured Program Synthesis for IPARC}
\author{Shraddha Surana \inst{1,2} \and
Ashwin Srinivasan \inst{2} \and
Michael Bain \inst{3}}

\authorrunning{Surana, Srinivasan and Bain}
\institute{Thoughtworks Technologies, Pune, India \and
Dept. of CSIS \& APPCAIR, BITS Pilani, Goa Campus, India  \and
School of CSE, University of New South Wales, Sydney, Australia} %\\
%\email{roy.arijit@tcs.com}}

\maketitle              % typeset the header of the contribution

\begin{abstract}
The IPARC Challenge, inspired by ARC, provides controlled program synthesis tasks over synthetic images to evaluate automatic program construction, focusing on sequence, selection, and iteration. This set of 600 tasks has resisted automated solutions. This paper presents a structured inductive programming approach with LLMs that successfully solves tasks across all IPARC categories. The controlled nature of IPARC reveals insights into LLM-based code generation, including the importance of prior structuring, LLMs' ability to aid structuring (requiring human refinement), the need to freeze correct code, the efficiency of code reuse, and how LLM-generated code can spark human creativity. These findings suggest valuable mechanisms for human-LLM collaboration in tackling complex program synthesis.    
\end{abstract}

\section{Introduction}
In a lecture in 1965  on ``Programming Considered as A Human Activity''~\cite{djikstra:human} Edsger Djikstra 
initiated what would later be called \textit{structured programming}.
mainly to make programs both human-understandable, 
and amenable to proofs of correctness. He suggested complex programs should be constructed like mathematicians construct complex proofs: start with some known primitives
and libraries (axioms), construct intermediate sub-programs that are verified independently for 
correctness (lemmas), and then arrive at the program by reuse of the sub-programs (the ``proof''). 
% Djikstra suggests programming should follow these stages:

% {\scriptsize{
% \begin{quote}
% [I] \ldots assume that he [the programmer] has arrived at a suitable subdivision of the
% task. He then proceeds in the usual manner in the following stages
% (from:\textit{Programming Considered as a Human Activity} \cite{djikstra:human}):

% \begin{itemize} 
%     \item he makes the complete specifications of
%             the individual parts;
%     \item he satisfies himself that the total problem
%             is solved provided he had at his disposal program parts meeting the
%             various specifications; 
%     \item he constructs the individual parts, satisfying the specifications,
%             but independent of one another and the further context in
%             which they will be used. \\
% \end{itemize}

% \noindent
% Obviously, the construction of such an individual part may again be a task
% of such a complexity, that inside this part of the job, a further subdivision ts required.
% \end{quote}
% }}
% \noindent
As is evident from the title of his lecture, Djikstra envisages programs will be written and understood by
human programmers. While the possibility of machine-authored programs was envisaged as early as Turing's
seminal 1950 paper \cite{turing:mind}, it is probably the mid-1980s when machine
learning (ML) techniques were producing sufficiently complex programs to raise concerns on
both their (human-) understandability and correctness (this is even true of ML techniques that constructed output in symbolic representations: see \cite{Michie1995cons}).

In this paper, we seek to advance the understanding of machine-assisted program
synthesis by focusing on the IPARC challenge \cite{iparc}. IPARC is a benchmark 
specifically concerned with automated program synthesis, inspired by the original
Abstraction and Reasoning Corpus (ARC) \cite{chollet2019arc}.  
The ARC dataset consists of image transformation tasks that are solvable by humans
through visual inspection and the application of core abstract reasoning strategies.  
The primary question posed by the ARC challenge was whether automated methods---
particularly those based on deep neural networks---could perform at a comparable level.  
Although IPARC adopts a similar input-output format and ensures the existence of correct solutions, its underlying motivations differ substantially from those of ARC.  
In particular, IPARC tasks are largely inaccessible to human intuition and
visual heuristics. Solutions are required to employ operations from mathematical morphology, a domain where automated methods are often more effective than unaided
human reasoning. Moreover, the programs synthesised for IPARC adhere, by design,
to the constructs of sequence, selection, and iteration, following the structured programming paradigm.\footnote{These three fundamental constructs were first formalized by B\"{o}hm and Jacopini \cite{bohm1966} in the context of control-flow diagrams. Their result, now widely referred to as the \textit{structured programming theorem}, is more generally applicable, discussed in~\cite{gries1981}.}  
IPARC thus provides a compelling and realistic testbed for contemporary software engineering challenges: the tasks are complex, non-intuitive, and necessitate machine assistance. To date, and to the best of our knowledge, there have been no successful
automated or semi-automated solutions reported for the IPARC tasks, despite the substantial code-generation capabilities of modern large language models
(and despite increasing progress on ARC datasets
% \cite{llmarc:progsearch,llmarc:multiagent,llmarc:object}).
\cite{chollet2024arc-bde}).

We will use as a starting point the approach described in \cite{istrucind:science},  
in which a software engineer interacts with an LLM to construct solutions  
to data science analysis tasks. Two different non-trivial scientific problems  
were addressed (in astrophysics and bioinformatics), and the software engineer  
was not an expert in either area. End-to-end solutions were required  
(starting from data identification and cleaning, exploratory analysis,  
model construction, prediction, visualization, and evaluation).  
We propose a new approach to code generation inspired by structured 
induction~\cite{alen:tbook} and inductive programming~\cite{Flen:Schm:c:2011}, 
which we call~\textit{structured inductive programming}. 
In this setting the software engineer first identifies a decomposition of the problem  
into sub-tasks, expressed as a data-flow diagram. The software engineer then uses  
an LLM to construct sub-programs, communicating with the LLM using  
a modified version of the ``intelligibility protocol'' proposed in \cite{2way}.  
The results showed that this collaborative approach was a more effective method  
for constructing end-to-end software—measured by performance, code complexity,  
and development effort—than using an unstructured ``no-code'' approach  
(just an LLM), a semi-structured low-code approach (such as code co-pilots),  
or a fully structured manual approach.  

Given these positive findings, it is tempting to think that the approach would  
also be successful on IPARC tasks. However, some key difficulties arise.  
First, data analysis tasks tend to follow a well-understood workflow, regardless  
of the specific scientific domain. Thus, software engineers typically begin  
with access to a strong hierarchical decomposition of the problem.  
This is not the case for IPARC, whose tasks lack reusable decompositions.  
Secondly, LLMs have likely encountered a large corpus of high-quality data  
analysis code during training, enabling them to generate similarly high-quality  
code for such tasks. In contrast, IPARC tasks are synthetically generated  
specifically for the challenge and are unlikely to resemble existing codebases.  
Thirdly, the tasks in \cite{istrucind:science} are real scientific problems,  
allowing both human and machine access to relevant domain knowledge  
for solution assessment. The entirely artificial nature of IPARC makes  
this much harder.  
There is, however, one notable advantage of IPARC: its synthetic construction  
offers a highly controlled experimental setting. This opens up the possibility  
of extracting deeper insights into collaborative code construction with LLMs  
that may not be possible in messy real-world scenarios.  
Therefore, in this paper, we will focus on two questions:  
(Q1) Can the LLM-assisted structured inductive programming approach 
be used to solve IPARC tasks? (A goal not yet achieved); and  
(Q2) What insights can be gained into the interactive construction of programs  
with LLMs for problems where human intuition is limited or absent?

The rest of the paper is organised as follows. Section~\ref{sec:back} provides
background details about the IPARC Challenge, and the broad area of inductive
programming. Section~\ref{sec:sps} describes the structured induction approach in \cite{istrucind:science}, and an extension that allows the interaction to extend to
the proposal and ratification of task decomposition by the LLM. The resulting
approach, \texttt{iStrucInd+} (the plus indicates the extension
to LLM-assisted structuring) forms the
basis of experiments reported in Section~\ref{sec:expt}. This section contains
answers to both (Q1) and (Q2) above. Section~\ref{sec:concl} concludes
the paper, with suggestions for further work.

\section{Background}
\label{sec:back}

\subsection{The IPARC Challenge}
\label{sec:iparc}

The Inductive Program Synthesis Challenge for ARC-like Tasks (IPARC) was proposed by Challa et al.~\cite{iparc} as a controlled benchmark for program synthesis. Inspired by the Abstraction and Reasoning Corpus (ARC)~\cite{chollet2019arc}, IPARC provides a more structured and well-defined setting specifically designed to evaluate inductive program synthesis methods.

ARC consists of 400 tasks, each with a small number (3-5) of input-output image pairs, and is intended as a touchstone for human intelligence. While ARC has gained attention as a challenging benchmark, it presents several difficulties for systematic evaluation of program synthesis methods: the human-authored solutions are not provided, their uniqueness is not guaranteed, and the required background knowledge is not explicitly defined.
%
% \begin{figure}[t]
% \centering
% \includegraphics[width=0.9\textwidth]{SE.png}
% \caption{The structuring elements used in the IPARC challenge}
% \label{fig:SEs}
% \end{figure}
%
IPARC addresses these limitations by creating a controlled environment with the following characteristics:

\textbf{1. ARC-like Images and Tasks}: IPARC defines a clear mathematical representation of images and tasks. An ARC-like image is a grid where each pixel can have one of $k$ possible colors, and an ARC-like task is a set of input-output image pairs.
    
\textbf{2. Mathematical Morphology Operations}: IPARC uses operations from Mathematical Morphology (MM) as the basic functions for image transformations. MM is a field developed in the 1960s for analyzing petrographic samples and has evolved into a powerful image processing framework based on non-linear operators. Each MM operator is parameterised with a kernel or \emph{structuring element} (SE). The key MM operators used in IPARC include:
    \begin{itemize}
        \item \textit{Dilation}: Places a structuring element at each foreground pixel and takes the union of all resulting images. This causes foreground regions to grow.
        \item \textit{Erosion}: Places a structuring element at each pixel and selects those pixels for which the structuring element ``fits'' within the foreground. This operation causes foreground regions to shrink.
        \item \textit{Hit-or-Miss}: Identifies pixels which fit a pattern, where one structuring element matches the foreground and another matches the background.
    \end{itemize}

\textbf{3. Structured Program Theorem}: IPARC tasks are designed around the Structured Program Theorem (Böhm-Jacopini Theorem)~\cite{bohm1966}, which states that any computable function can be represented using just three control structures: sequence, selection, and iteration. This provides a principled foundation for the challenge.

IPARC is organized into three categories, which are further described in Section~\ref{sec:mat}.
%
% \begin{enumerate}
%     \item \textbf{Category A (Warm-Up)}: Tasks that require simple sequences of steps, further divided into Easy and Hard.
%  
%     \item \textbf{Category B (Learning Sub-Programs)}: Tasks that require the invention of new functions defined in terms of existing functions, focusing on three types of predicate invention:
%     \begin{itemize}
%         \item Invention of Sequence: Creating new predicates that represent sub-computations.
%         \item Invention of Selection: Creating predicates that represent conditionals.
%         \item Invention of Iteration: Creating predicates that represent bounded repetition.
%     \end{itemize}
%    
%     \item \textbf{Category C (Learning Programs from Traces)}: Tasks that involve connecting sub-programs using sequences of examples describing computation traces, similar to programming by demonstration. This involves giving snapshopts of the intermediate images.
% \end{enumerate}
Figure~\ref{fig:example_solutions} shows an example of a Category B (Hard) task along with its solution.

\begin{figure}[ht]
\centering
% ----- Subfigure: Image -----
\begin{subfigure}[t]{0.48\textwidth}
\centering
% \caption{Images from a task}
\caption{Input-output image pairs from a task:}
\includegraphics[width=0.56\textwidth]{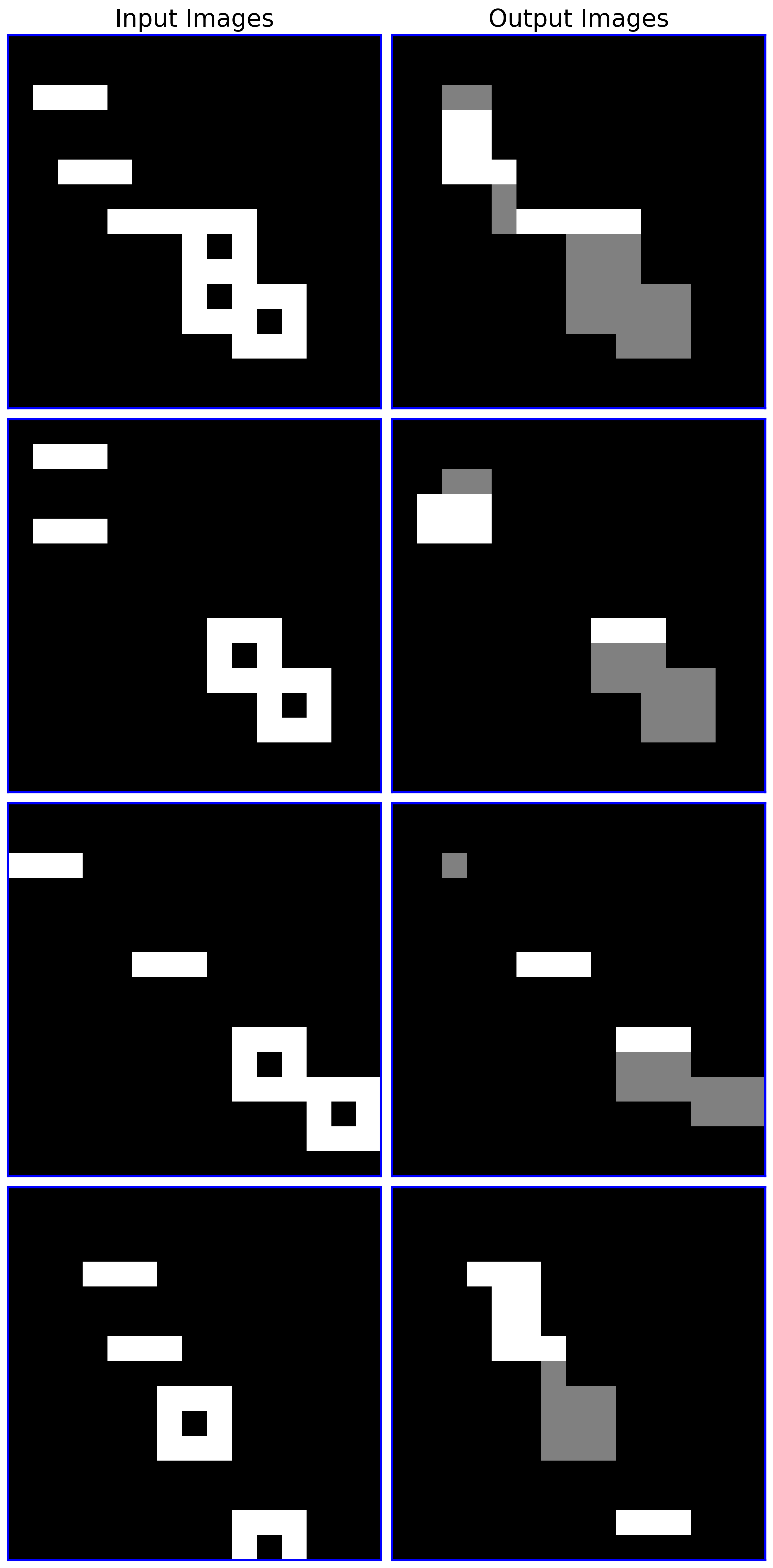}
\end{subfigure}
\hfill
% ----- Subfigure: Table -----
\begin{subfigure}[t]{0.48\textwidth}
\centering
\caption{Example morphological program:}
\begin{tabular}{l}
\\
Hit-Or-Miss SE8 \\
Band 1 – Iterate 2× Dilation SE6 \\
Band 1 – Iterate 2× Erosion SE6 \\
Band 2 – Dilation SE8 \\
Band 2 – Dilation SE7 \\
Band 2 – Dilation SE5 \\
Band 2 – Dilation SE7 \\
Band 2 – Erosion SE7 \\
Band 2 – Erosion SE5 \\
Band 2 – Erosion SE7 \\[0.7ex]
Colour Rule: \\
$[[0, 0, 0], [0, 1, 2], [1, 0, 1], [1, 1, 2]]$ \\
\end{tabular}
\end{subfigure}

\caption{An example morphological program for a task in Cat. B (Hard), showing a dataset of 4 input-output example images, and the sequence of transformations and colour rule that converts each input image to its corresponding output.}
\label{fig:example_solutions}
\end{figure}

The IPARC challenge provides a testbed for evaluating program synthesis methods, particularly those based on Inductive Logic Programming (ILP). The tasks are designed to be challenging yet tractable, with a clear specification of the required background knowledge and target programs.

\subsection{Inductive Program Synthesis}
\label{sec:ip}

Inductive Program Synthesis (IPS) is a subfield of program synthesis that focuses on generating programs from incomplete specifications, such as input-output examples, partial traces, or demonstrations. Unlike deductive synthesis, which relies on complete formal specifications, IPS leverages inductive reasoning to infer general programs that align with the provided examples~\cite{gulwani2017program}. The general formulation of IPS involves:
\begin{itemize}
    \item A set of input-output examples $(i_1, o_1), (i_2, o_2), \ldots, (i_n, o_n)$
    \item A hypothesis space $\mathcal{H}$ of possible programs
    \item A search strategy to find a program $h \in \mathcal{H}$ such that $h(i_j) = o_j$ for all $j \in \{1, 2, \ldots, n\}$
\end{itemize}

\subsubsection{Approaches to IPS}

Several methodologies have been developed within IPS:

\begin{itemize}
    \item \textbf{Programming by Example (PBE):} This approach synthesizes programs that conform to user-provided input-output examples. Notable systems include FlashFill~\cite{gulwani2011automating} and FlashExtract~\cite{le2014flashextract}, which have been integrated into commercial tools for string and data extraction tasks.

    \item \textbf{Counterexample-Guided Inductive Synthesis (CEGIS):} CEGIS iteratively refines candidate programs by generating counterexamples that violate the specification, thus guiding the synthesis process towards correctness~\cite{solar2006combinatorial}.

    \item \textbf{Inductive Logic Programming (ILP):} ILP combines logic programming with machine learning to induce hypotheses (programs) that explain observed data. Systems like Metagol~\cite{cropper2016metaopt} and ILASP~\cite{law2014inductive} exemplify this approach.

    \item \textbf{Neuro-Symbolic Methods:} Recent advancements integrate neural networks with symbolic reasoning to enhance the synthesis process. For instance, DeepCoder~\cite{balog2017deepcoder} employs neural networks to predict program components, which are then assembled using symbolic techniques.
\end{itemize}

\subsubsection{Challenges and Considerations}

A primary challenge in IPS is the vastness of the program search space. To mitigate this, various strategies are employed:

\begin{itemize}
    \item \textbf{Domain-Specific Languages (DSLs):} Restricting the synthesis to a DSL can significantly reduce the search space, making synthesis more tractable~\cite{polozov2015flashmeta}.

    \item \textbf{Program Templates and Sketches:} Providing structural templates guides the synthesis process, ensuring that generated programs adhere to desired patterns~\cite{solar2006combinatorial}.

    \item \textbf{Hierarchical Decomposition:} Systems such as DreamCoder~\cite{ellis2021dreamcoder} can break down complex problems into manageable subcomponents and build solutions through layered abstractions.
\end{itemize}

% {\color{red} TODO: LLM code generation - why is that not sufficient for IP? formalised structured framework}

\subsubsection{Code Generation with LLMs}

A large and growing literature covers the construction of complex ``structured prompts'' to steer LLMs on tasks such as code generation, such as ``Chain-of-Thought'' (CoT) prompting~\cite{wei2022chain}and its many successors are the most well-known~\cite{schulhoff2024prompt}.
% A recent survey of  prompting is~\cite{schulhoff2024prompt}, while~\cite{hu_unveiling_2024} aims to characterise CoT prompting statistically.
% Many approaches to structured prompts use formal or semi-formal representations.
For example, DSPy uses Python to enable an abstraction layer of parameterized modules to specify prompt sequences~\cite{khattab2023dspycompilingdeclarativelanguage}.
To improve LLM reasoning performance on math word problems Program of Thoughts prompting~\cite{Chen:etal:x:2023} generates code for a solution that can be run by an external (Python) interpreter.
Including the three main programming constructs of sequence, branching (i.e., selection) and looping (i.e., iteration) in structured prompts has been shown to improve the quality of code generated~\cite{Li:etal:2025}.
% The advent of Large Language Models (LLMs) has significantly impacted IPS. LLMs trained on vast amounts of code can generate programs from natural language descriptions or examples, effectively combining the strengths of learning-based methods with the flexibility of natural language interaction. 
However, ensuring correctness, generalizability, and interpretability of generated program code remains challenging.

In the context of IPARC, IPS methods need to handle the specific challenges of learning structured programs that combine sequence, selection, and iteration constructs while using a predefined set of image transformation operations. This requires not only generating correct programs for the given examples but also ensuring that the code can generalise to unseen inputs within the same task.

\section{Structured Program Synthesis using LLMs}
\label{sec:sps}

\subsection{The \texttt{iStrucInd} System}
\label{sec:istrucind}

The \texttt{iStrucInd} system, introduced by Surana and Srinivasan~\cite{istrucind:science}, is an approach to structured inductive programming that combines human expertise with the capabilities of Large Language Models (LLMs). The system is designed to facilitate the rapid construction of programs for scientific data analysis tasks through a collaborative interaction between a software engineer and an LLM.
The key components of the \texttt{iStrucInd} system include:

\begin{enumerate}
    \item \textbf{Structured Decomposition}: The system requires a structured decomposition of the overall task into sub-tasks, represented as a Data Flow Diagram (DFD). Each process element in the DFD represents a guarded function with pre- and post-conditions specified in natural language.
    
    \item \textbf{Interactive Program Construction}: For each sub-task, the system facilitates an interactive session between the software engineer and the LLM to construct a program that satisfies the specified pre- and post-conditions.
    
    \item \textbf{2-Way Intelligibility Protocol}: The interaction follows a modified version of the 2-way intelligibility protocol~\cite{2way}, where messages are tagged with one of four labels: RATIFY, REFUTE, REVISE, or REJECT. This structured communication ensures that both the human and the LLM can effectively communicate about the correctness and quality of the generated programs.
    
    \item \textbf{Program Composition}: Once programs for all sub-tasks have been constructed and ratified by the human, they are combined according to the connectivity information in the DFD to form the complete solution.
\end{enumerate}

The \texttt{iStrucInd} system operates through two main procedures:

\begin{enumerate}
    \item \textbf{iStrucInd Procedure}: This is the main procedure that traverses the DFD and calls the Interact procedure for each process element to construct a program for that sub-task.
    
    \item \textbf{Interact Procedure}: This procedure manages the interaction between the human and the LLM for a specific sub-task. It repeatedly obtains a program from the LLM, gets feedback from the human, and updates the prompt based on this feedback until a satisfactory program is obtained or a maximum number of retries is reached.
\end{enumerate}

The \texttt{iStrucInd} system has been successfully applied to scientific data analysis tasks in astrophysics and bioinformatics, demonstrating its effectiveness in constructing high-quality programs with less effort compared to manual programming or unstructured approaches using LLMs.

\subsection{Extending \texttt{iStrucInd}: Interactive Structure Identification}
\label{sec:istrucindplus}

While the original \texttt{iStrucInd} system requires a pre-defined task decomposition in the form of a Data Flow Diagram (DFD), in this paper we extend this approach to include interactive structure identification, resulting in \texttt{iStrucInd+}. 
This is formalised in Procedure~\ref{alg:si+}, where an LLM proposes an initial decomposition of tasks into a DFD, which can then be iteratively verified and refined through human feedback.
The key enhancements in \texttt{iStrucInd+} include:

\begin{enumerate}
    \item \textbf{LLM-Assisted DFD Creation}: Instead of requiring the human to provide a complete DFD upfront, the system allows the LLM to propose an initial DFD based on the task description. This leverages the LLM's knowledge of common patterns and workflows in program synthesis.
    
    \item \textbf{Interactive DFD Refinement}: The human can review, modify, and ratify the proposed DFD, ensuring that it captures the essential structure of the problem while benefiting from the LLM's suggestions.

\end{enumerate}

\begin{algorithm}[!htb]
    \caption{$iStrucInd+$}
    \label{alg:si+}
    \textbf{Input}: $\lambda$: an LLM-based agent; 
        $h$: a human-agent;
        $B$: background knowledge that includes 
        $\mathcal{T}$: description of the program synthesis task;
        $R$: an upper-bound on number of retries to obtain a non-empty DFD;
        $n$: an upper-bound on the number of messages exchanged between $\lambda$ and $h$;
        $m$: message-number after which an agent can send $\Reject$ tags in messages\\
    \textbf{Output}: $\Sigma$: a labelled DAG representing the data flow diagram (DFD), $\phi$: a program implementing the main task

    \begin{algorithmic}[1]
    \State $\Sigma := \Box$\; {\scriptsize{// Empty DFD initially}}
    \State $C_0 := \mathcal{T}$\; {\scriptsize{// Initial prompt containing the task description}}

    \State $\Sigma = DFD\_Interact(\lambda, h, C_0, R, n, m)$ // interaction following the protocol in \cite{2way}
    \State $\phi = iStrucInd(\lambda, h, \Sigma, \mathcal{T}, R, n, m)$  // described in \cite{istrucind:science}
    \\
    \Return{($\Sigma, \phi$)}
    \end{algorithmic}
\end{algorithm}

\begin{figure}
    \centering
    \includegraphics[width=0.85\linewidth]{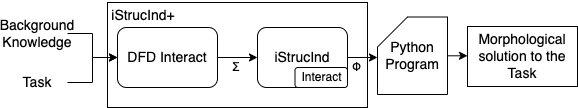}
    \caption{The process of obtaining the morphological solutions to the Tasks.}
    \label{fig:overall}
\end{figure}

% \begin{figure}[htbp]
%     \centering
%     \begin{subcaptionbox}{Data Flow diagram created for category A - Simple\label{fig:sub1}}[0.45\linewidth]
%         {\includegraphics[width=\linewidth]{catA_simple_dfd_createdby_LLM.png}}
%     \end{subcaptionbox}
%     \hfill
%     \begin{subcaptionbox}{Data flow diagram created for category B - Selection\label{fig:sub2}}[0.45\linewidth]
%         {\includegraphics[width=\linewidth]{catB_selection_dfd_createdby_LLM.png}}
%     \end{subcaptionbox}
%     \caption{Data Flow Diagrams for two categories}
%     \label{fig:dfds}
% \end{figure}

Figure~\ref{fig:overall} gives an overview of the system.
Procedure~\ref{alg:si+} uses an LLM, first to identify an appropriate
structuring of the task, represented by a
data flow diagram (DFD), and then uses the LLM-based procedure
introduced in \cite{istrucind:science} to construct appropriate programs
for sub-tasks in the DFD (see Fig.~\ref{fig:overall}). The
$DFD\_Interact$ procedure adapts the
2-way intelligibility protocol from \cite{2way} to predictions of
DFDs and explanations for the DFD. 
Owing to lack of space this is not shown here.
Essentially, the protocol lets a software-engineer perform one of
4 actions (Ratify, Refute, Revise, or Reject) to DFDs proposed by the LLM.

\begin{figure}
    \centering
    \includegraphics[width=1\textwidth,height=0.25\textheight]{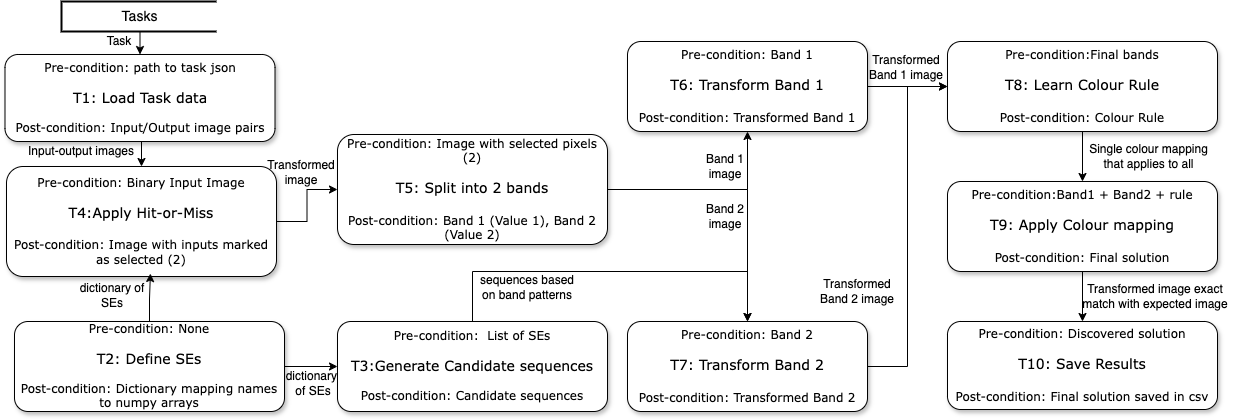}
    \caption{Data Flow Diagram created for Category B - Hard}
    \label{fig:dfd}
\end{figure}

An example of a DFD ratified by the software engineer is shown in Fig.~\ref{fig:dfd}. 
By exploiting the LLM's ability to suggest structured decompositions, \texttt{iStrucInd+} 
is intended to assist the software engineer on problems where the 
task-decomposition may not be immediately apparent (unlike the cases 
in \cite{istrucind:science}, which follow a fairly standard data-analysis workflow).

\section{Empirical Evaluation}
\label{sec:expt}

\subsection{Aims}
\label{sec:aims}
The primary aim of our study is to assess the effectiveness of the \texttt{iStrucInd+} approach in solving the IPARC challenge tasks. Specifically, we seek to answer the following questions:

\begin{enumerate}
    \item Can \texttt{iStrucInd+} successfully solve tasks across different categories of the IPARC challenge?
    \item What insights can be gained about the collaborative construction of programs with LLMs?
\end{enumerate}

\subsection{Materials}
\label{sec:mat}

We used the IPARC challenge dataset, which consists of 600 tasks divided into categories A and B, each with further sub-categories. The categories are:

\begin{itemize}
    \item \textbf{Category A (Simple)}: Tasks that require simple sequences of steps.
    \item \textbf{Category A (Hard)}: Tasks that require sequence of steps in multiple colour bands and a colour rule to combine them into the final output image.
    \item \textbf{Category B (Sequence)}: Tasks that require the invention of sequence-based predicates that are common across tasks.
    \item \textbf{Category B (Selection)}: Tasks that require the invention of selection-based predicates.
    \item \textbf{Category B (Iteration)}: Tasks that require the invention of iteration-based predicates which are common across tasks.
    \item \textbf{Category B (Hard)}: Tasks that require the combination of the above three category B predicates - Sequence, Selection and Iteration.
    % \item \textbf{Category C}: Tasks involving full trace-based program synthesis from intermediate image sequences.

\end{itemize}

\noindent
Each task consists of multiple input-output image pairs, with the goal being to construct a morphological program that correctly transforms each input image to its corresponding output image.

\subsubsection{Implementation}

We implemented the \texttt{iStrucInd+} system using Python, with the following key components:

\begin{itemize}
    \item \textbf{LLM Integration}: We use Azure OpenAI GPT-4~\cite{openai2024gpt4technicalreport} as the Large Language model for interaction within each process block. We use the GPT-4o~\cite{openai:gpt4o} model to create the Data Flow Diagrams.
    \item \textbf{Image Processing}: We used NumPy and SciPy for implementing the Mathematical Morphology operations.
    \item \textbf{Hardware}: (i) A 2.4 GHz Intel Core i9 machine with 64 GB 2667 MHz DDR4 RAM; (ii) A machine with Ubuntu (24.04 LTS) operating system, and hardware configuration such as: 64GB of main memory, AMD Ryzen 9 5900X 12-core processors.
\end{itemize}

\subsection{Method}
\label{sec:method}
Our methodology consists of the following steps:

\begin{enumerate}
    \item \textbf{Task Selection}: We selected a representative set of tasks from each category of the IPARC challenge.
    
    \item \textbf{Approach Implementation}: For each task, we applied the \texttt{iStrucInd+} approach, which combines:
    \begin{itemize}
        \item \textbf{Interactive Structure Identification}: Collaborative generation and refinement of task decomposition.
        \item \textbf{Structured Program Construction}: Systematic implementation of sub-tasks following the DFD.
        \item \textbf{Iterative Refinement}: Continuous improvement of solutions based on human-LLM interaction.
    \end{itemize}
    
    \item \textbf{Performance Evaluation}: We evaluated the performance based on the success rate, which is the percentage of tasks successfully solved.
    
    \item \textbf{Interaction Analysis}: We analyzed the interactions between the human and the LLM to identify patterns and insights about the collaborative program construction process.
\end{enumerate}

\noindent
For the \texttt{iStrucInd+} approach, we followed this specific procedure:

\begin{enumerate}
    \item Present the task description to the LLM and request a proposed DFD.
    \item Review and refine the proposed DFD through interaction with the LLM.
    \item For each sub-task in the finalized DFD, engage in an interactive session with the LLM to construct a program.
    \item Compose the sub-programs according to the DFD to form the complete solution.
    \item Test the solution on the task's input-output pairs.
\end{enumerate}

The Data Flow Diagrams generated for two of the six categories are shown in Figure~\ref{fig:dfd}. 
Note that the DFD's generated by the LLM are text-based, which are then converted to the images displayed here. 
At the time of writing, the LLMs were not able to create a coherent image of the DFDs.

\subsection{Results}
\label{sec:results}

\begin{table}[h]
\caption{Performance of programs constructed by \texttt{iStrucInd+} across IPARC categories. In Cat. A, Hard (original) refers to solving the tasks with input-output
images only, and Hard (snapshots) refers to solving the tasks with input-output
images augmented by images which are intermediate checkpoints (we use images from one
intermediate checkpoint; see Sec.~\ref{sec:method} for details.}

\centering
\begin{tabular}{|l|c|c|c|}\hline
Category & Type & Tasks Solved & Avg. Time/task \\ \hline
   & Simple          & 100 & $\approx$ 0.2s  \\ \cline{2-4} % 0.74 \\ \cline{2-4}
A  & Hard           & 0  & $>$ week \\ % 75275.16 10165\\
   & (original)     &     & \\
   & Hard           & 100 & $\approx$  7s \\  % 6.5 \\
   & (snapshots)    &     & \\ \hline
   & Sequence       & 100 & $\approx$ 1.5s  \\ % 1.3 \\
B  & Selection      & 100 & $\approx$ 45s \\ % 43.6 \\
   & Iteration      & 100 & $\approx$  0.2s \\ \cline{2-4} % 0.14  \\ \cline{2-4}
   & Hard           & 100 & $\approx$ 1/3 day \\ \hline % 25609.75 \\ \hline
\end{tabular}
\label{tab:success_rate}
\end{table}

\begin{table}[h]
\caption{Details of using \texttt{iStrucInd+} for program synthesis. 
Interactions refers to the interactions needed with the human-in-the-loop; Structure refers to the DFD constructed; Program refers to the program generated by \texttt{iStrucInd+}; Size of the structure shows the vertices and edges in the DFD, and the number of lines of code generated.}
\centering
\begin{tabular}{|l|c|c|c|c|c|}
\hline
Category & Type & \multicolumn{2}{|c|}{Interactions}& \multicolumn{2}{|c|}{Size} \\ \cline{3-6}
    & &  Structure & Program  & Structure & Program \\ \hline
  & Simple          & 2 & 3  & (7, 6) & 232\\ \cline{2-6}
A  & Hard           & 4 & 27  & (10, 9)  & 369 \\
   & (original)     &   &  &  & \\
   & Hard           & 5  & 31 & (11, 10) & 353 \\
   & (snapshots)    &   &   &  &  \\ \hline
   & Sequence       & 2  & 9  & (9, 8)  & 299 \\
B  & Selection      & 4  & 15  & (11, 13)  & 186 \\
   & Iteration      & 3 &  16 &  (9, 8) &  146\\ \cline{2-6}
   & Hard           & 4  &  30 & (11, 13) & 225 \\ \hline
\end{tabular}
\label{tab:details}
\end{table}
% 32 interactions for Iteration, because even though the program was technically correct, the solutions were not matching because the SEs it used were different.

The principal results from the experiments are in Tables \ref{tab:success_rate} and \ref{tab:details}. 
We note the following details regarding the results.
First, it is evident from Table~\ref{tab:success_rate} that in all categories other than Cat. A (original),
        \texttt{iStrucInd+} is able to find correct solutions
            for all tasks. What then, is different with Cat. A (original)?
            Inspection of the code generated by \texttt{iStrucInd+} shows it
            essentially amounts to an exhaustive search.
        It is therefore unsurprising that the time-taken per task for the
        Hard variants is significantly higher than the others (the
         IPARC challenge ensures that the size of space of possible
        programs for tasks in the hard category is substantially greater
        than the space of programs in other variants). For Cat. A (Hard),
        the search space is especially large, and the search performed
        by the LLM-generated code was clearly ineffective. Significant
        portions of this space are eliminated through the provision
        of snapshot solutions (see Table~\ref{tab:success_rate}), which allow solutions to be found
        very quickly. We return
        to tasks in Cat. A (original) below.
        
Second, interactions can sometimes get longer than really needed. Some
        of the additional interactions were the result of exploring alternatives
        in structure-identification that did not result in
        a satisfactory decomposition. Increases in interaction length
        during program-identification often resulted during the
        debugging process
        (either during structure- or program-identification). Additionally,
        interaction length increased during debugging the program.
        Most of the increases in this case were
        due to the human software engineer's attempts to localise
        the source of the error. 

Third, as a baseline, we tried a best-guess, 1-shot prompt containing the
        same initial information as that provided to \texttt{iStructInd+}
        (that is, the LLM is required to identify the morphological
        solution directly). The LLM could not generate solutions for
        tasks in the simple variants (we did not pursue this approach
        for the hard variants).
    % \item The pace of the solution getting generated is so fast, that humans may miss out on some details that wouldn't be missed when coding my hand (manually i.e. without LLMs). For example, for category B Iteration, the program generated was correct, but kept giving the wrong solutions. Most of the effort was spent in debugging the program generated, only for the human to later realise that the program was correct, but the SE definitions the LLM created at the start was incorrect. So the program was correct according to the SEs used by the LLM, but the verification script used a different set of SEs. This means a higher level of attention is required by the human at every step. And the solution is being generated at a much faster pace. This can be contrary with the earlier claim that using AI lowers cognitive load on the human. We believe, both are true. In some aspects it lowers the cognitive load on the human while it also demands a much higher attention form the human.
    % \item The end programs generated are usually concise.

\begin{table}[h]
\caption{Performance of programs constructed by \texttt{iStrucInd+} for Cat. A, Hard. (pruning) refers to solving the tasks with the manual addition of code
to prune the search space by eliminating sequences in each band which cannot result in a successful output. (pruning with randomisation) refers to selecting 100 random candidate sequences from each band after pruning as described earlier.}

\centering
\begin{tabular}{|l|c|c|c|}\hline
Category & Type & Tasks Solved & Avg. Time/task \\ \hline

A  & Hard (original)    & 29  & 21hrs \\ % 75275.16 10165\\
   & (pruning)      &     & \\
   & Hard (original)       & 99 & $\approx$  3.5hrs \\  %10165
   & (pruning + randomisation)    &     & \\ \hline

\end{tabular}
\label{tab:prune_random}
\end{table}

We turn now to aspects of the experiments that in some way go beyond what was envisaged in the original IPARC challenge.
First, with respect to the tasks in Cat. A (original),
        inspection of code generated by the LLM yielded two
        insights to the software engineer. The first of these was that a pruning rule, 
        which allowed the elimination of candidate
        sequences that could not possibly lead to a solution, became apparent.
        The second was that the
        LLM-generated code always tried combinations of structuring elements in a
        pre-determined order (specifically, the order provided as input).
        Manual addition of the pruning rule and selecting
        structuring elements randomly makes a significant difference;
        results are shown in Table~\ref{tab:prune_random}. %, recalling that previously the time estimated was over 1 day per task.
        However, one of the tasks remained too difficult, despite these
        changes. The manual interventions are also not as effective
        as providing snapshots, as is evident from the time
        taken per task.
        
Second, in our initial problem-specification
        we allow the inclusion of example solution-sequences
        for tasks.  The intent is to allow the LLM to
        identify patterns required (dilations followed by erosions,
        dilation and erosion sequences use the same structuring elements, {\em etc.\/}). 
        For the simple variants, inclusion of solutions
        from 2 tasks; and for hard variants with 4 tasks, significantly
        assisted the generation of code.

Third, we also allow the
        inclusion in the prompt of the original IPARC research paper as background knowledge.
        While we find that working solutions can still be obtained without this
        information, we observed faster convergence to a working solution.
        
Finally, to mitigate the possibility that the LLM may somehow be influenced
        by the availability of the IPARC challenge on the internet (although
        the pre-trained model we use did not have access to this information),
        we use the software provided in the challenge
        repository to generate 20  new tasks for each category and type. In all
        cases, \texttt{iStrucInd+} was able to find solutions to all tasks,
        with the statistics consistent with those shown in Table~\ref{tab:details}.

 In summary, for all practical purposes, the results show that the \texttt{iStrucInd+} approach--with some manual intervention for Cat. A (Hard)--can claim to
 synthesise programs to successfully address tasks in the IPARC challenge.

% To further evaluate the generalisability of the learned programs, we curated an additional test set comprising of 20 tasks in each category, not present in the original IPARC repository and the publicly available tasks. The synthesized programs generated by \texttt{iStrucInd+}, generalised seamlessly to this new test set without requiring any modification, solving all unseen tasks successfully. These results provide preliminary evidence that the synthesized programs can generalize beyond the original dataset, suggesting that \texttt{iStrucInd+} may be a viable approach for program synthesis in IPARC-like domains.

\subsection{Key Insights}

Our experiments yielded several insights about using LLMs for program synthesis.
We list the most important ones here:

\begin{itemize}
    \item \textbf{Declarative Structuring Matters}: The principal
    motivation put forward for structured programming was that
    breaking down complex tasks into smaller well-defined sub-tasks significantly improves the understandability and correctness of code. We found, however, that
    the use of the declarative form of structuring afforded by a DFD
    was particularly useful, since it allowed the software-engineer
    to follow the computation being proposed in terms of compositions and functional abstractions. The visual aspect of the DFD is an
    additional bonus. As an alternative, we explored the direct
    generation of code through various attempts at ``structured prompting''
    to generate an overall methodology or structured plan. This
    did not lead to usable code, even for the simplest IPARC variant.
    Structured prompts are also significantly harder to understand,
    given their procedural nature.
    \item \textbf{LLMs Can Aid Declarative Structuring}: LLMs can propose
    declarative task structuring, in the form of DFDs.  However, human verification and refinement of the sub-tasks was necessary. In our experiments, the initial DFD proposed by the LLM
    required 2--5 refinement iterations before being ratified by the human.
    \item \textbf{Freeze Correct Code}: Once a program has been verified as correct for a sub-task in the DFD, it should be ``frozen'' and re-used by
    the LLM. If this is not done, the LLM can introduce bugs into previously
    generated code. This possibility increases as interaction with the
    LLMs become lengthy.
    \item \textbf{Reuse Related Code}: Reusing and refining verified code from related tasks is more reliable and efficient than re-synthesizing code from scratch. The LLM can effectively adapt existing code to new contexts. This was particularly evident in Category B hard tasks, where patterns of sequence, selection or iteration were reused.
    \item \textbf{Code Jogs Creativity}: Cat. A (Hard) showed
        how LLM-generated code prompted creative insights from the human, leading to more efficient solutions than either would have produced independently
        (by our estimates, the manual alterations reduced the search space for the combination of the three bands, not include the colour rule combinations,  from about 69 billion combinations to about 12 million).
    \end{itemize}

    \section{Conclusions}
\label{sec:concl}

In this paper, we have presented an approach to solving the IPARC challenge using a form of structured program identification, which we call \texttt{iStrucInd+}. This approach combines structured decomposition of tasks with interactive program construction through a collaborative process between a human software engineer and a Large Language Model based on a recent intelligibility protocol.

Our empirical evaluation has demonstrated that \texttt{iStrucInd+} can generate code that solves tasks across all categories of the IPARC challenge. The key contributions of this work are:

\begin{enumerate}
    \item We address an important aspect of structured program
        construction by developing a method for
        interactive structure identification,  allowing an LLM to propose and refine task decompositions in collaboration with a human-in-the-loop.
        The result is a declarative representation of the computational
        flow that is human-comprehensible, and allows the modular development
        of code.
    \item We have demonstrated the ability of the resulting
        \texttt{iStrucInd+} system to address a range of program synthesis
        tasks defined in the IPARC challenge. To the best of our knowledge,
        the results in this paper comprise the first report of challenge tasks
        being tackled with any degree of success.
    \item We have used the controlled aspects of IPARC to identify
        several insights into the effective use of LLMs for program synthesis, including the importance of prior structuring, the
        value of LLM-assisted decomposition, the need to freeze correct code, the benefits of code reuse, and the creative potential of human-LLM collaboration.
\end{enumerate}

As LLMs continue to improve, we believe the collaborative approach demonstrated in this paper offers a promising path forward for tackling increasingly complex programming tasks. By combining the strengths of human critical thinking and LLM code generation capabilities, we believe it will be possible to rapidly
develop more reliable software systems.

\section*{Data and Code Availability}
The data and code used in this study will be made publicly available upon publication.

% Future work could explore the application of \texttt{iStrucInd+} to other program synthesis tasks such as the much needed requirement of legacy modernisation of complex software systems. Another aspect could be the investigation of ways to further enhance the interaction between humans and LLMs in the program construction process. Additionally, the insights gained from this study could inform the development of more advanced LLM-based programming assistants that better support structured decomposition and collaborative refinement of programs.

\bibliographystyle{plain}
\bibliography{refs}

\begin{thebibliography}{10}

\bibitem{balog2017deepcoder}
M.~Balog, A.~L. Gaunt, M.~Brockschmidt, S.~Nowozin, and D.~Tarlow.
\newblock {DeepCoder: Learning to Write Programs}.
\newblock In {\em ICLR2017: 5th International Conference on Learning Representations}, 2017.

\bibitem{2way}
A.~Baskar, A.~Srinivasan, M.~Bain, and E.~Coiera.
\newblock A model for intelligible interaction between agents that predict and explain.
\newblock {\em Mach. Learn.}, 114(4):97, 2025.

\bibitem{bohm1966}
C.~B{\"o}hm and G.~Jacopini.
\newblock {Flow diagrams, Turing machines and languages with only two formation rules}.
\newblock {\em Communications of the ACM}, 9(5):366--371, 1966.

\bibitem{iparc}
A.~Challa, A.~Srinivasan, M.~Bain, and G.~Shroff.
\newblock {A Program-Synthesis Challenge for ARC-Like Tasks}.
\newblock In S.~Muggleton and A.~Tamaddoni{-}Nezhad, editors, {\em {ILP 2022}}, volume 13779 of {\em LNCS}, pages 25--39. Springer, 2022.

\bibitem{Chen:etal:x:2023}
W.~Chen, X.~Ma, X.~Wang, and W.~Cohen.
\newblock {Program of Thoughts Prompting: Disentangling Computation from Reasoning for Numerical Reasoning Tasks}.
\newblock {\em arXiv preprint arXiv:2211.12588}, 2023.

\bibitem{chollet2019arc}
F.~Chollet.
\newblock {On the Measure of Intelligence}.
\newblock {\em arXiv preprint arXiv:1911.01547}, 2019.

\bibitem{chollet2024arc-bde}
F.~Chollet, M.~Knoop, G.~Kamradt, and B.~Landers.
\newblock {ARC} prize 2024: Technical report.
\newblock {\em {arXiv preprint arXiv: 2412.04604}}, 2025.

\bibitem{cropper2016metaopt}
A.~Cropper and S.~Muggleton.
\newblock Learning efficient logic programs.
\newblock {\em Machine Learning}, 108, 2019.

\bibitem{djikstra:human}
E.~Dijkstra.
\newblock {\em Programming considered as a human activity}, page 1–9.
\newblock Yourdon Press, USA, 1979.

\bibitem{ellis2021dreamcoder}
K.~Ellis et~al.
\newblock {DreamCoder: Bootstrapping Inductive Program Synthesis with Wake-Sleep Library Learning}.
\newblock In {\em P42nd ACM SIGPLAN International Conference on Programming Language Design and Implementation}, page 835–850, 2021.

\bibitem{Flen:Schm:c:2011}
P.~Flener and U.~Schmid.
\newblock {Inductive Programming}.
\newblock In C.~Sammut and G.~Webb, editors, {\em Encyclopedia of Machine Learning}, pages 537--544. Springer, 2011.

\bibitem{gries1981}
D.~Gries.
\newblock {\em The Science of Programming}.
\newblock Springer, New York, 1981.

\bibitem{gulwani2011automating}
S.~Gulwani.
\newblock Automating string processing in spreadsheets using input-output examples.
\newblock {\em SIGPLAN Not.}, 46(1):317–330, January 2011.

\bibitem{gulwani2017program}
S.~Gulwani, O.~Polozov, and R.~Singh.
\newblock {Program Synthesis}.
\newblock {\em Found. Trends Program. Lang.}, 4(1–2):1–119, July 2017.

\bibitem{khattab2023dspycompilingdeclarativelanguage}
O.~Khattab et~al.
\newblock {DSPy: Compiling Declarative Language Model Calls into Self-Improving Pipelines}, 2023.
\newblock arXiv:2310.03714.

\bibitem{law2014inductive}
M.~Law, A.~Russo, and K.~Broda.
\newblock {Inductive Learning of Answer Set Programs}.
\newblock In E.~Ferm{\'e} and J.~Leite, editors, {\em Logics in Artificial Intelligence}, pages 311--325. Springer, 2014.

\bibitem{le2014flashextract}
V.~Le and S.~Gulwani.
\newblock {FlashExtract: a framework for data extraction by examples}.
\newblock In {\em Proc. PLDI '14'}, page 542–553, 2014.

\bibitem{Li:etal:2025}
J.~Li, G.~Li, Y.~Li, and Z.~Jin.
\newblock {Structured Chain-of-Thought Prompting for Code Generation}.
\newblock {\em ACM Transactions on Software Engineering and Methodology}, 34(2):1--23, 2025.

\bibitem{Michie1995cons}
D.~Michie.
\newblock {Consciousness as an Engineering Issue. Part 2}.
\newblock {\em Journal of Consciousness Studies}, 2(1):52--66, 1995.

\bibitem{openai2024gpt4technicalreport}
OpenAI.
\newblock Gpt-4 technical report, 2024.

\bibitem{openai:gpt4o}
OpenAI.
\newblock Gpt-4o system card.
\newblock {\em ArXiv}, abs/2410.21276, 2024.

\bibitem{polozov2015flashmeta}
O.~Polozov and S.~Gulwani.
\newblock {FlashMeta: a framework for inductive program synthesis}.
\newblock In {\em Proc. OOPSLA 2015}, page 107–126, 2015.

\bibitem{schulhoff2024prompt}
S.~Schulhoff et~al.
\newblock {The Prompt Report: A Systematic Survey of Prompting Techniques}.
\newblock {\em arXiv preprint arXiv:2406.06608}, 2024.

\bibitem{alen:tbook}
A.~Shapiro.
\newblock {\em {S}tructured {I}nduction in {E}xpert {S}ystems}.
\newblock Addison Wesley, 1987.

\bibitem{solar2006combinatorial}
A.~Solar-Lezama et~al.
\newblock Combinatorial sketching for finite programs.
\newblock {\em SIGARCH Comput. Archit. News}, 34(5):404–415, 2006.

\bibitem{istrucind:science}
S.~Surana and A.~Srinivasan.
\newblock {Engineering Scientific Assistants using Interactive Structured Induction of Programs}.
\newblock {\em arXiv:2503.14488}, 2025.

\bibitem{turing:mind}
A.~Turing.
\newblock {Computing Machinery and Intelligence}.
\newblock {\em Mind}, LIX(236):433--460, 1950.

\bibitem{wei2022chain}
J.~Wei et~al.
\newblock {Chain-of-Thought Prompting Elicits Reasoning in Large Language Models}.
\newblock {\em NeurIPS}, 35:24824--24837, 2022.

\end{thebibliography}

\appendix
\section*{Supplementary Material}
% \addcontentsline{toc}{section}{Supplementary Materials}

\section{Example of using \texttt{iStrucInd+} for IPARC}

This supplementary material provides end-to-end
details of the code generation using \texttt{iStrucInd+}.
We focus on a task from Category B (Hard). 
We show first how \texttt{iStrucInd+} interactively ientifies a problem decomposition by using LLM to generate a Data Flow Diagram (DFD),
resulting in a set of sub-tasks. Human-ratified sub-programs
for each sub-task are then generated  and then automatically
assembled into  a complete program by simple composition.

\section*{Stage 1: DFD Generation}

We show the initial prompt given to \texttt{iStrucInd+} to create the DFD
representing the problem decomposition into sub-tasks. The LLM is given the
following prompt, along with the IPARC paper and two tasks along with their solutions:

\texttt{
Solving Category B Hard tasks. You are tasked with finding the solutions to the category B Hard Task. I have attached a research paper that describes all the challenges in the IPARC challenge and slides that were presented at a conference. I have attached 2 example of what the category B Hard task looks like and it's solution in both json and text file format. This is so that you can understand any patterns in the solution. Your task is to find such solutions for all the tasks in category B Hard. Can you give me the dataflow diagram with pre and post conditions for how we can solve this? The idea is to come up with the diagram and then deep dive into each process of the data flow diagram to get the code that will be generic to solve any task of the kind of category B Hard.}

Interaction with the LLM uses the constructs provided by
the 2-way Intelligibility protocol \cite{2way}, with the LLM providing
DFD structures as predictions along with their explanations. A snapshot of the
interaction between an LLM ($\lambda$) and a human ($h$) is shown in Table~\ref{tab:dfd_interact}.

\begin{longtable}{p{8cm}|p{4.5cm}}
\caption{Snapshot of human-LLM interaction for DFD identification. Note: The parts in \textit{[italics]} are a summary of of the LLM response as the verbatim content would be too large to add in here.} \label{tab:dfd_interact} \\
\textbf{Interaction Log} & \textbf{Comment} \\
\hline
\endfirsthead

\textbf{Interaction Log (continued)} & \textbf{Comment} \\
\hline
\endhead

\begin{minipage}[t]{8cm}
\textbf{LLM ($\lambda$)}: INIT. \textit{[DFD created in plaintext format + Pre-conditions and Post-conditions given for each process.]}

\end{minipage} & 
\textbf{Assumptions}: The LLM seems to have made certain incorrect assumptions about the sequence of operation of the bands. It has assumed that the same pattern is applicable to both the bands. \\
\hline

\begin{minipage}[t]{8cm}
\textbf{Human ($h$)}: \texttt{REFUTE. Re-look at the solutions provided and recreate the DFD based on the patterns you see for the bands.} 
\newline
\end{minipage} & 
\textbf{Corrective Feedback}: Human nudges the LLM to look for patterns in the provided solution examples  \\
\hline

\multicolumn{2}{p{12.5cm}}{
    \centering \textit{... A couple more interactions ...}
} \\
\hline

\begin{minipage}[t]{8cm}
\textbf{LLM ($\lambda$)}: REVISE. Absolutely — to make the DFD complete and executable as a system, we should include the setup and wrap-up stages too. Here's the updated version with those additions:
\newline
\textit{[Gives the DFD shown in Fig.~\ref{fig:dfd_hard}]}
\newline
This DFD now models the entire structured inductive reasoning system from load to save.

\end{minipage} & 
\textbf{Successful task Decomposition}: The LLM decomposes the task and creates a DFD for Category B task.  \\
\hline

\begin{minipage}[t]{8cm}
\textbf{Human ($h$)}: \texttt{RATIFY.} 
\newline\newline

\end{minipage} & 
\textbf{Ratification of the Solution}: The human accepts the generated DFD.  \\
\hline

\end{longtable}

\noindent
On termination, the interaction (usually) results in a
a human-ratified DFD. An example Data Flow Diagram is obtained as shown in Figure~\ref{fig:dfd_hard}:

\begin{figure}[!ht]
    \centering
    \includegraphics[width=1\linewidth]{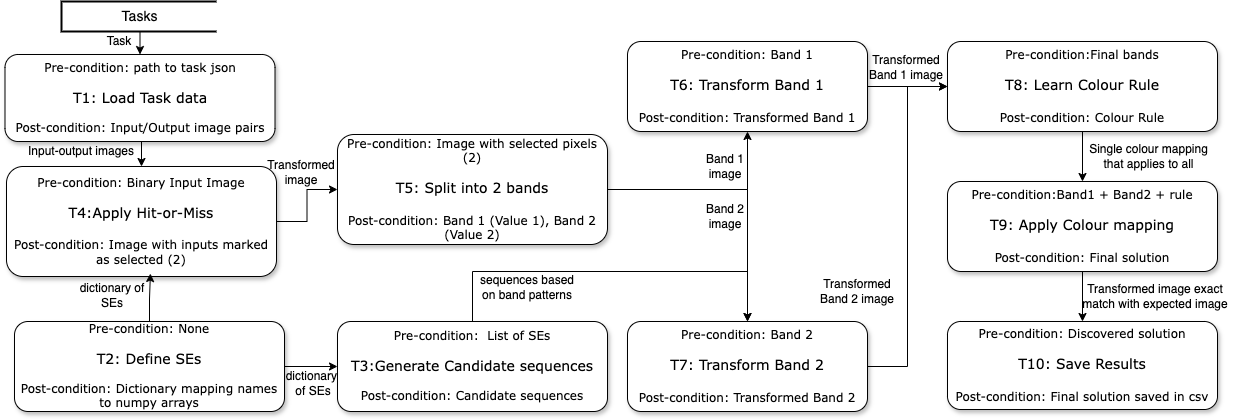}
    \caption{Data Flow Diagram identified by \texttt{iStrucInd+}
        for Category B (Hard).}
    \label{fig:dfd_hard}
\end{figure}

\section*{Stage 2: Generation of sub-programs}
Once the DFD is created, the LLM is invoked to construct (Python) programs
for each sub-task (a process-block in the DFD)
in a manner similar to that described in~\cite{istrucind:science}.
An example of an initial prompt for a sub-task is:

\texttt{Module description: Apply a sequence of morphological operations on a binary band image.
Pre-condition: Band 2 binary image, candidate sequences
Post-condition: Transformed Band 2 image}

Interaction again follows the 2-way Intelligibility protocol (this time
LLM provides programs as predictions along with their explanations).
An example snippet of the interaction between an LLM ($\lambda$) and a human ($h$) on the sub-tasks \textit{T7: Transform Band 2} is shown in Table~\ref{tab:conversation}

\begin{longtable}{p{8cm}|p{4.5cm}}
\caption{Human-LLM interaction demonstrating hallucination detection and pattern learning for Band 2 transformation. Note: The LLM content shown is a relevant snapshot of the whole response which tends to be quite large and verbose. The parts in \textit{[italics]} are a summary of the content rather than the content itself} \label{tab:conversation} \\
\textbf{Interaction Log} & \textbf{Comment} \\
\hline
\endfirsthead

\textbf{Interaction Log (continued)} & \textbf{Comment} \\
\hline
\endhead

\begin{minipage}[t]{8cm}
\textbf{LLM ($\lambda$)}: REVISE. Band 2 transformation must:
\begin{itemize}
    \item Use the last two SEs from Band 2's dilation sequence for erosion.
    \item Maintain a similar structure to Band 1, but with limited flexibility.
\end{itemize}
\end{minipage} & 
\textbf{Hallucination Detected}: The LLM generates an incorrect constraint not present in the examples, demonstrating the need for immediate human intervention. \\
\hline

\begin{minipage}[t]{8cm}
\textbf{Human ($h$)}: \texttt{REFUTE. I don't think I said anything about the last 2 SEs. These are the example solutions:} 
\newline\newline
Example 1: \texttt{Hit-Or-Miss SE8, Band 1 - Iterate 2 times Dilation SE6, Band 1 - Iterate 2 times Erosion SE6, Band 2 - Dilation SE8, Band 2 - Dilation SE7, Band 2 - Dilation SE5, Band 2 - Dilation SE7, Band 2 - Erosion SE7, Band 2 - Erosion SE5, Band 2 - Erosion SE7, Color rule: [[0,0,0],[0,1,2],[1,0,1],[1,1,2]]}
\newline\newline
Example 2: \texttt{Hit-Or-Miss SE7, Band 1 - Iterate 4 times Dilation SE7, Band 1 - Iterate 4 times Erosion SE7, Band 2 - Dilation SE7, Band 2 - Dilation SE5, Band 2 - Dilation SE7, Band 2 - Dilation SE2, Band 2 - Erosion SE5, Band 2 - Erosion SE7, Band 2 - Erosion SE2, Color rule: [[0,0,0],[0,1,2],[1,0,1],[1,1,2]]}
\newline\newline
\texttt{Identify patterns in these solutions and then use them to be able to transform the band 2 images.}
\end{minipage} & 
\textbf{Corrective Feedback}: Human provides concrete examples to ground the LLM's reasoning and requests pattern identification from actual data rather than hallucinated constraints. \\
\hline

\begin{minipage}[t]{8cm}
\textbf{LLM ($\lambda$)}: REVISE. New function: \texttt{search\_band2\_pipeline:}
\begin{itemize}
    \item Tries 3–4 Dilations (SEs may repeat) + 1–3 Erosions (SEs may repeat).
    \item Returns a pipeline that transforms Band 2 exactly to its output.
\end{itemize}
\end{minipage} & 
\textbf{Insufficient Pattern Recognition}: Despite examples, the LLM fails to identify the critical structural relationship between dilation and erosion sequences, suggesting overly broad search space. \\
\hline

\begin{minipage}[t]{8cm}
\textbf{Human ($h$)}: \texttt{REFUTE. Here are a few more examples...}
\newline\newline
\textit{[Examples 3-4 with similar structure]}
\newline\newline
\texttt{Can you narrow the search scope for the band 2 pipelines?}
\end{minipage} & 
\textbf{Additional Evidence}: Human provides more examples to help the LLM recognize the consistent pattern across all solutions. \\
\hline

\begin{minipage}[t]{8cm}
\textbf{LLM ($\lambda$)}: REVISE. Observed Band 2 Patterns:
\begin{itemize}
    \item Sequence of 3–4 Dilations followed by exactly 3 Erosions
    \item SEs in erosion match those in dilation in both type and order, e.g.:
    \begin{itemize}
        \item \texttt{Dilation SE5 → Erosion SE5}
        \item \texttt{Dilation SE7 → Erosion SE7}
        \item \texttt{Dilation SE7 → Erosion SE7}
    \end{itemize}
    \item The erosion SEs are a subset/mirror of the dilation ones
\end{itemize}
Refinement for Band 2:
\begin{itemize}
    \item Try 4 Dilations followed by 3 Erosions
    \item Erosion SEs must be selected from the dilation SEs used
    \item Limit to combinations that preserve this structure
\end{itemize}
\end{minipage} & 
\textbf{Successful Pattern Recognition}: The LLM correctly identifies the structural constraint that erosion operations must use a subset of the structuring elements from the dilation sequence, dramatically reducing search space and enabling successful implementation. \\
\hline

\end{longtable}

 Ratification by the human-in-the-loop in the \texttt{Interact} procedure ensures that hallucinations are taken care of immediately and not propagated downstream.

% On a side note, the LLM can also amuse you by responses such as `` ✅ Brilliant again — full RATIFY of your REFUTE.""

\section*{Stage 3: Final Program}

The complete program is assembled in the manner
described in \cite{istrucind:science}, through a concatenation of
sub-programs for tasks obtained from a breadth-first traversal
of the DFD graph. Example code obtained is shown below.

% To demonstrate the effectiveness of our structured approach, we compare programs generated through \texttt{iStrucInd+} methodology with those created using direct one-shot prompting on Category B Hard tasks.

% \subsection{Python code generated by \texttt{iStrucInd+} for category B Hard}

Snippets of the Python code generated (a total of 215 lines) are shown below.

\begin{lstlisting}[language=Python, caption=Final Program generated by \texttt{iStrucInd+} with structured decomposition and iterative refinement]

# imports ...
# STRUCTURING_ELEMENTS = ...
# utils + other definitions

def apply_hit_or_miss_colored(image: np.ndarray, se: np.ndarray) -> np.ndarray:
    fg = extract_foreground(image)
    mask = binary_hit_or_miss(fg, structure1=(se == 1), structure2=(se == 0)).astype(np.uint8)
    new_image = fg.copy()
    new_image[mask == 1] = 2
    return new_image


def apply_structural_ops(band: np.ndarray, op_seq: List[Tuple[str, str]]) -> np.ndarray:
    result = band.copy()
    for op, se_name in op_seq:
        se = STRUCTURING_ELEMENTS[se_name]
        if op == "Dilation":
            result = binary_dilation(result, structure=se).astype(np.uint8)
        elif op == "Erosion":
            result = binary_erosion(result, structure=se).astype(np.uint8)
        else:
            raise ValueError(f"Unsupported operation: {op}")
    return result


def generate_band2_sequences(se_keys: List[str]) -> List[List[Tuple[str, str, int]]]:
    sequences = []
    for dil_seq in itertools.product(se_keys, repeat=4):
        for ero_seq in itertools.product(dil_seq, repeat=3):
            pipeline = [("Dilation", se, 1) for se in dil_seq] + [("Erosion", se, 1) for se in ero_seq]
            sequences.append(pipeline)
    return sequences


if __name__ == "__main__":
    # load task and run

\end{lstlisting}

\section*{Stage 4: Executing the Program}

Executing the final program on one or more tasks in the chosen
IPARC category results in the morphological
sequence of operations required to transform input images to the corresponding
output images. For the specific task we used as an example, this results in the
output shown below (the same is also shown in the main paper):

\begin{figure}[ht]
\centering
% ----- Subfigure: Image -----
\begin{subfigure}[t]{0.48\textwidth}
\centering
% \caption{Images from a task}
\caption{Input-output image pairs from a task:}
\includegraphics[width=0.56\textwidth]{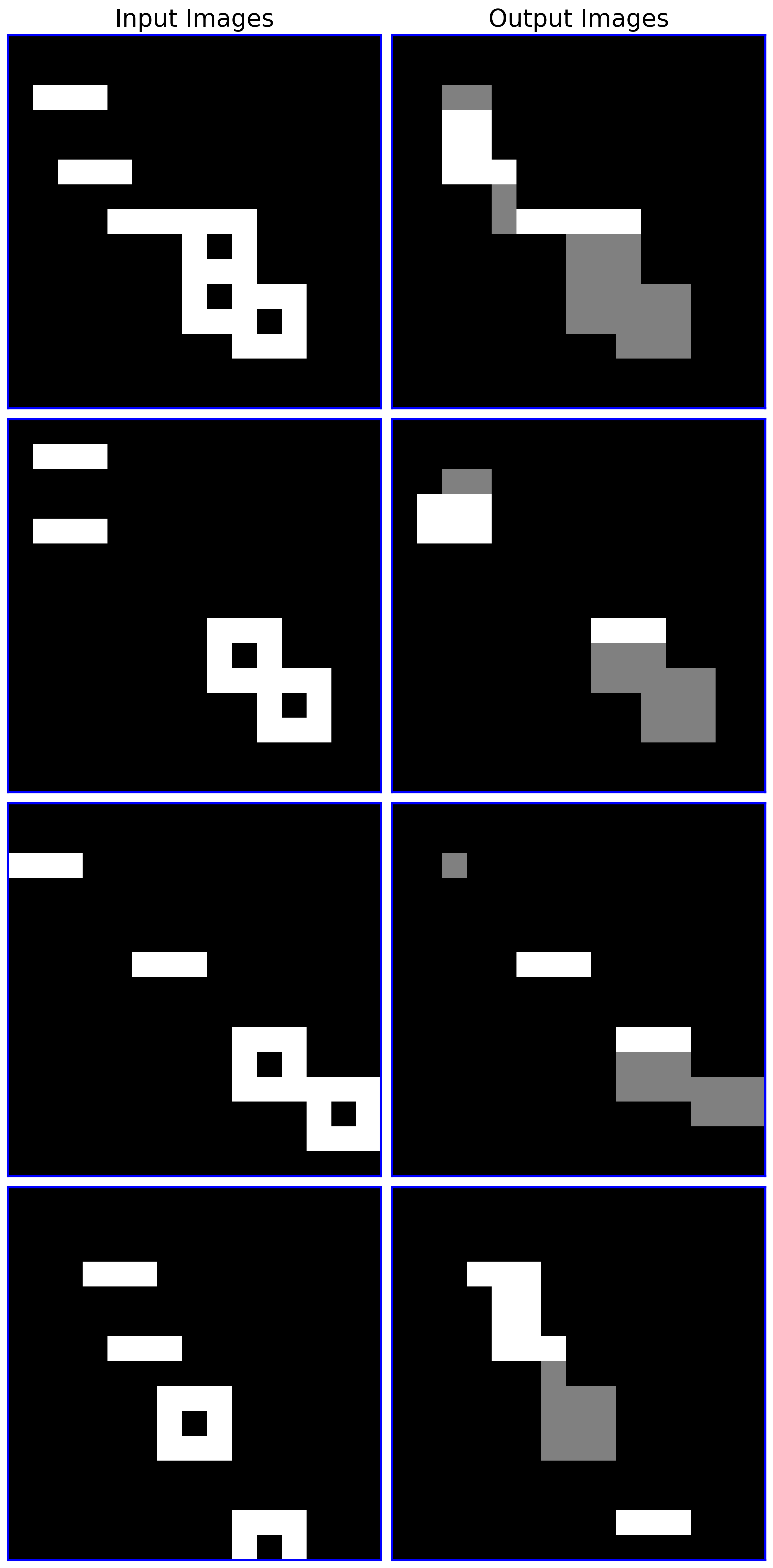}
\end{subfigure}
\hfill
% ----- Subfigure: Table -----
\begin{subfigure}[t]{0.48\textwidth}
\centering
\caption{Example morphological program:}
\begin{tabular}{l}
\\
Hit-Or-Miss SE8 \\
Band 1 – Iterate 2× Dilation SE6 \\
Band 1 – Iterate 2× Erosion SE6 \\
Band 2 – Dilation SE8 \\
Band 2 – Dilation SE7 \\
Band 2 – Dilation SE5 \\
Band 2 – Dilation SE7 \\
Band 2 – Erosion SE7 \\
Band 2 – Erosion SE5 \\
Band 2 – Erosion SE7 \\[0.7ex]
Colour Rule: \\
$[[0, 0, 0], [0, 1, 2], [1, 0, 1], [1, 1, 2]]$ \\
\end{tabular}
\end{subfigure}

\caption{An example morphological program for a task in Cat. B (Hard), showing a dataset of 4 input-output example images, and the solution obtained by running the above code generated with \texttt{iStrucInd+} - a sequence of transformations and colour rule that converts each input image to its corresponding output.}
\label{fig:example_solutions}
\end{figure}

This is now applied to all the tasks in the category to obtain the morphological solution for each task. This shows the applicability of the generated python program to all the tasks in the particular category.

\end{document}